\documentclass[preprint,english,prb,superscriptaddress,showpacs,amsmath,amssymb]{revtex4}
\usepackage{graphicx,esint,bm,float,lipsum,babel}
\makeatletter
\@ifundefined{textcolor}{}
{%
 \definecolor{BLACK}{gray}{0}
 \definecolor{WHITE}{gray}{1}
 \definecolor{RED}{rgb}{1,0,0}
 \definecolor{GREEN}{rgb}{0,1,0}
 \definecolor{BLUE}{rgb}{0,0,1}
 \definecolor{CYAN}{cmyk}{1,0,0,0}
 \definecolor{MAGENTA}{cmyk}{0,1,0,0}
 \definecolor{YELLOW}{cmyk}{0,0,1,0}
 }
\makeatother
\def\Re{{\rm Re}}
\def\cond{{\rm c}}
\def\val{{\rm v}}
\def\kvec{{\bf k}}
\def\kpoint{{\rm K}}
\def\kprime{{\rm K^{\prime}}}
\def\rvec{{\bf r}}
\def\Rvec{{\bf R}}
\def\qvec{{\bf q}}
\def\kpoint{{\rm K}}
\def\kprime{{\rm K^{\prime}}}
\def\elaser{E_{\rm L}}

\def\qbwf{1/q_{\rm BWF}}
\def\ers{{\rm ERS}}

\def\efermi{E_{\rm F}}
\def\kdir{K^{\rm d}}

\begin{document}

\title{Breit-Wigner-Fano lineshapes in Raman spectra of graphene}

\author{Eddwi~H.~Hasdeo} \email [Electronic Address: ]
{hasdeo@flex.phys.tohoku.ac.jp} \affiliation{Department of Physics,
  Tohoku University, Sendai 980-8578, Japan}
\author{Ahmad~R.~T.~Nugraha} \affiliation{Department of Physics,
  Tohoku University, Sendai 980-8578, Japan}
\author{Mildred~S.~Dresselhaus} \affiliation{Department of Electrical
  Engineering and Computer Science, Massachusetts Institute of
  Technology, Cambridge, Massachusetts 02139-4037, USA}
\affiliation{Department of Physics, Massachusetts Institute of
  Technology, Cambridge, Massachusetts 02139-4307, USA}
\author{Riichiro~Saito} \affiliation{Department of Physics, Tohoku
  University, Sendai 980-8578, Japan}

\date{\today}

\begin{abstract}
  Excitation of electron-hole pairs in the vicinity of the Dirac
  cone by the Coulomb interaction gives rise to an asymmetric
  Breit-Wigner-Fano lineshape in the phonon Raman spectra in graphene.
  This asymmetric lineshape appears due to the interference effect
  between the phonon spectra and the electron-hole pair excitation
  spectra. The calculated Breit-Wigner-Fano asymmetric factor
  $1/q_{\rm BWF}$ as a function of the Fermi energy shows a ``V''-shaped
  curve with a minimum value at the charge neutrality point and gives
  good agreement with the experimental result.
\end{abstract}


\maketitle
\section{Introduction}
Elementary excitations such as electrons and phonons can be probed by
the inelastic scattering of light using the Raman spectroscopy
technique.  In graphene-related systems, studying the shape of the
Raman spectra can give us a deep understanding of the electron energy
dispersion,~\cite{malard07} phonon energy dispersion,~\cite{mafra07}
lifetime of excitations,~\cite{lazzeri06} the Kohn anomaly
effect,~\cite{piscanec04} and structure
characterization.~\cite{dresselhaus10} In particular, the asymmetric
Breit-Wigner-Fano (BWF) lineshape, historically observed in the Raman
spectra of graphite interacalation compounds (GICs)~\cite{eklund79}
and metallic nanotubes (m-SWNTs),~\cite{brown01} probes interference
between the continuum spectra with discrete spectra.~\cite{fano61}
Recently, the BWF asymmetry has been observed by Yoon~\emph{et
  al.}~\cite{yoon13} in monolayer graphene indicating a common origin
of the BWF lineshape of the graphite-related systems (i.e. GICs,
m-SWNTS, and monolayer graphene) that arise due to the presence of the
Dirac cone or linear energy band structure.

The BWF lineshape is defined by the following formula
\begin{align}
\label{eq:ibwf}
I_{\rm BWF}(\omega_{\rm s})=&I_0 \frac{(1+s/q_{\rm BWF})^2}{1+s^2}\nonumber\\
=&I_0 \left[\frac{1}{q_{\rm BWF}^2}+ \frac{1-1/q_{\rm
      BWF}^2}{1+s^2}+\frac{2s/q_{\rm BWF}}{1+s^2}\right],
\end{align} 
where $s=(\omega_{\rm s}-\omega_{\rm G})/\Gamma$.  Here $\omega_{\rm
  s}$, $\omega_{\rm G}$, $\qbwf$, $\Gamma$, and $I_0$ are the Raman
shift, the spectral peak position, the asymmetric factor, the spectral
width, and the maximum intensity of the BWF spectra, respectively. The
right hand side of Eq.~\eqref{eq:ibwf} tells us that the BWF lineshape
respectively consists of a constant continuum spectrum, a discrete
Lorentzian spectrum, and an interference effect between both
spectra. When $\qbwf=0$, Eq.~\eqref{eq:ibwf} gives a Lorentzian
lineshape which represents a discrete phonon spectrum. The
interference term gives rise to an asymmetric lineshape for positive
and negative values of $s$, in which the asymmetry is proportional to
a dimensionless parameter $\qbwf$, mimicking the ratio between the
probability amplitude of the continuum spectra to that of the discrete
spectra.~\cite{fano61}

In the Raman spectroscopy studies of graphite-related systems,~\cite{jorio11,
  saito11} continuum spectra come from the electronic excitations and
are usually observed only in metallic systems.  The BWF lineshapes in
graphene have been found in various kinds of phenomena such as
scanning tunneling microscopy,~\cite{wehling10} optical
conductivity,~\cite{mak11} photoabsorption spectroscopy,~\cite{chae11}
and infrared spectroscopy~\cite{kuzmenko09,tang10} revealing that
electron-hole pair excitations in the vicinity of the Dirac cone play an
important role in the continuum spectra.

The asymmetric BWF lineshapes in graphite-related systems are normally
found in the Raman shift around $1600\ \rm cm^{-1}$, known as the $G$
modes, which correspond to two zone-center ($\qvec=0$) phonon modes,
namely the in-plane tangential optic (iTO) and longitudinal optic
modes. In graphene, the BWF asymmetry of the $G$ band is observed
using the gate-modulated Raman spectroscopy.~\cite{yoon13} The
asymmetric factor ($\qbwf$) has a value around $-0.06$ or one-order of
magnitude smaller than those found in m-SWNTs ($\qbwf \approx -0.4
$)~\cite{brown01} and in GICs ($\qbwf \approx -0.5$).~\cite{eklund79}
The absolute value of the BWF asymmetric factor greatly decreases as
we change the Fermi energy ($\efermi$) to be further from the Dirac
point by applying a positive or a negative bias with respect to the
charge neutrality point.~\cite{yoon13,saito13} These results give a
clue that the asymmetric factor strongly depends on the electronic
density of states (DOS) near the Dirac cone.
\begin{figure*}[t]
\includegraphics[clip,width=16cm]{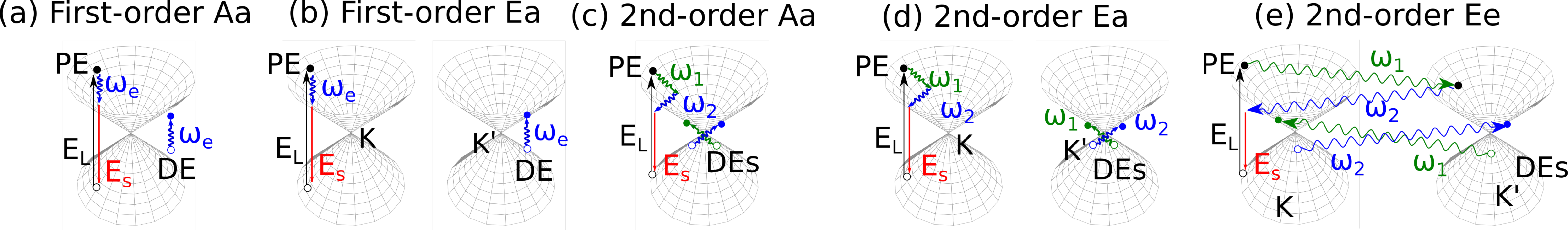}
\caption{\label{Fig1}(Color online) All possible Coulomb interactions
  between a photoexcited electron (PE) and electrons on the Dirac cone
  (DEs): (a) a first-order intravalley interaction and intravalley
  scattering (Aa), (b) a first-order intervalley interaction and
  intravalley scattering (Ea), (c) a second-order intravalley
  interaction and intravalley scattering (Aa), (d) a second order
  intervalley interaction and intravalley scattering (Ea), and (e) a
  second order intervalley interaction and intervalley scattering
  (Ee).  Capital (small) letters A \{a\} and E \{e\} label the
  intravalley and the intervalley interactions \{scatterings\}. The
  Raman shift $\omega_{\rm s}$ is the energy difference between laser
  excitation energy $\elaser$ and the scattered photon energy $E_{\rm
    s}$ which corresponds to the energy used to excite a DE (DEs) in
  the first(second)-order processes, $\omega_{\rm s} = \omega_{\rm e}(
  \omega_{\rm s} =\omega_1+\omega_2)$. }
\end{figure*}

In this work, we show that the origin of the BWF spectra in graphene
comes from the continuous single particle electron-hole pair spectra,
interfering with the discrete phonon spectra.  Hereafter, we refer to
the single particle electron-hole pair spectra as the electronic Raman
spectra (ERS).~\cite{farhat11} In the previous work for m-SWNTs, we
discuss that the ERS spectra originate from the second-order Coulomb
interaction with non-zero momentum transfer $\qvec \ne 0$, due to the
symmetry of the A and B sublattice wavefunctions which gives rise to the
absence of the direct Coulomb interaction at the zone center
$\qvec=0$.~\cite{hasdeo13} Unlike the previous calculation for m-SWNTs
which utilized exciton wavefunctions,~\cite{hasdeo13} in this
calculation we use electron wavefunctions from the tight binding (TB)
method because our calculation regime ($2.4$ eV) is far from the
saddle point energy dispersion ($4$ eV), and thus the exciton effects
are negligible.~\cite{chae11,yang09} The use of electron wavefunctions
give considerable contributions of the intervalley scattering to the
Raman intensity which was neglected in the previous study.\cite{hasdeo13}
After calculating the Raman amplitudes of the ERS and the phonon spectra, we
found that the interference between the ERS and the phonon spectra
gives a drastic change in the  constructive-destructive interference near
the phonon spectra, giving an asymmetry to the phonon lineshape when fitted
to the BWF lineshape.  By considering the second-order Raman process,
we systematically reproduce the $\efermi$ dependence of the Raman spectra
of graphene that was observed in experiment.~\cite{yoon13}

Our paper is organized as follows. In Sec.~\ref{sec:theor} we describe
our calculation of the electron-electron interaction using the TB
method and considering up to second-order Raman processes.  In
Sec.~\ref{sec:ers}, we discuss the calculated ERS spectra as a
function of $\efermi$ and compare the asymmetric BWF factor $\qbwf$
obtained from our calculation with that from the experiment.
Finally, in Sec.~\ref{sec:summ} we give a summary of this work.

\section{Theoretical methods}
\label{sec:theor}
The possible ERS processes are described in Fig.~\ref{Fig1}, which
consist of either intravalley (A) or intervalley (E) interaction,
either intravalley (a) or intervalley (e) scattering and either zero
momentum transfer ($\qvec= 0$ first-order) or non-zero momentum
transfer ($\qvec\ne 0$ second-order) processes.~\cite{hasdeo13} When a
photon with the laser excitation energy $\elaser$ is introduced to the
graphene sample, the photon excites an electron from an initial state
$i$ to an intermediate state $n$ with an energy matched to $\elaser$
(incident resonance). This photoexcited electron (PE) is then
scattered to another intermediate state $n'$ by the Dirac electrons
(DEs) on the Dirac cone and the electron finally recombines with a
hole by emitting a scattered photon energy $E_{\rm s}$ as shown in
Fig.~\ref{Fig1}. The Coulomb interaction between the PE and the DEs
causes the PE to reduce its energy and changes PE's momentum while the
DEs are being excited.

The number of DEs to be excited for each process depends on the number
of the scattering order.  In the first-order process, only one DE is
excited and this process requires a zero momentum transfer ($\qvec =
0$) since the PE momentum ($\kvec$) should be the same as its hole
momentum in order to emit a scattered photon with energy $E_{\rm s}$
by the electron-hole recombination process. In the second-order
process, on the other hand, the PE is scattered twice ($\kvec
\rightarrow \kvec-\qvec$) and ($\kvec-\qvec \rightarrow \kvec$) and
the PE excites two DEs with relative non-zero electron-hole momenta
$-\qvec$ and $\qvec$. Due to the degeneracy of the Dirac cone at the
$\kpoint$ and $\kprime$ points of the graphene Brillouin zone (BZ),
both the first-order processes and the second-order processes may
occur in the intravalley (A) interactions or in the intervalley (E)
interactions. In the A interactions, the DEs are excited on the same
Dirac cone as the PE, while in the E interactions, the DEs are excited
on the other Dirac cone.  In the case of the E interaction, the
initial and final states of the PE and DEs can be in the same
(different) valley which is defined by intravalley (intervalley)
scattering labelled by a small letter ``a'' (``e''). The e scattering is
not possible in the A interaction because $+\qvec$ and $-\qvec$
scattering are pointing to two different directions at the high
symmetry points of graphene; one is pointing to the $\overline{\rm
  KK^{'}}$ direction while the other is pointing to the $\overline{\rm
  K\Gamma}$ direction. Thus the Ae interaction does not conserve energy
during the scattering processes. Combining all possible A and E
interactions with the a and e scatterings we have: an Aa
[Fig.~\ref{Fig1}(a)] and an Ea [Fig.~\ref{Fig1}(b)] in the first-order
processes; and an Aa [Fig.~\ref{Fig1}(c)], an Ea [Fig.~\ref{Fig1}(d)],
and an Ee [Fig.~\ref{Fig1}(e)] in the second-order processes.

The BWF asymmetry comes from the interference effect mentioned above
because both the ERS and the phonon spectra have the same initial and
final states for a single PE. Based on this standpoint, we define the
Raman intensity:
\begin{equation}
  \label{eq:intens}
  I(\omega_{\rm s} )= \left[A_G(\omega_{\rm s} )+A_{\rm ERS}(\omega_{\rm s} )\right]^2,
\end {equation}
where $A_G=\sum_\nu A_{\nu}$, in which $A_\nu$ and $A_\ers$ are, respectively,
the $\nu$-th phonon scattering amplitude and the ERS scattering
amplitude. The phonon scattering amplitude in the resonance Raman
spectra is given by~\cite{yu10b}
\begin{align}
  A_{\nu}(\omega_{\mathrm{s}}) = & \frac {1}{\pi} \sum_{n,n^\prime}
  \bigg[ \frac{{\cal M}_{\rm el-op}^{n,i}} {\left[\Delta
      E_{ni}-i\gamma\right]}\nonumber\\& \times \frac{{\cal M}_{{\rm
      el-}\nu}^{n^\prime,n}} {\left[\Delta E_{n^\prime
        i}-\hbar\omega_{G}-i(\gamma+\Gamma_{\nu})\right]}\nonumber\\&
  \times \frac{{\cal M}_{\rm el-op}^{f,n^\prime}}
  {\left[E_{\mathrm{L}}-\hbar\omega_{G}-\hbar \omega_{\rm s}
      -i\Gamma_{\nu}\right]} \bigg], \label{eq:amplitude}
\end{align}
where for the phonon modes we only consider the first-order process
$\nu$ = iTO or LO modes, and $\Delta E_{ni}=E_{\rm L}-E_{n}-E_{i}$.
Here we use a broadening factor $\gamma = 0.1$ eV, which is related to
the inverse of the life time of the photoexcited carriers.  On the
other hand, $\Gamma_\nu$ is related to the life time of the
electron-phonon interaction.~\cite{sato10}

The values of $\Gamma_\nu$ and $\omega_{\rm G}$ are considered as
follows. In the gate-modulated Raman spectra, we expect phonon
frequency softening and spectral broadening as we shift the Fermi
energy from $|\efermi|>0$ to $\efermi = 0$.  This effect is due to the
Kohn anomaly, i.e. renormalization of the phonon energy by
electron-hole pair excitation in the $G$ mode Raman spectra of
graphene.~\cite{piscanec04} We do not consider the Kohn anomaly effect
explicitly in this calculation, but we can fit the peak position
$\omega_{\rm G}=1591+15|\efermi|\ {\rm cm^{-1}}$ for $-0.20\le \efermi
\le 0.00$ eV and $\omega_{\rm G}=1591+22.5|\efermi|\ {\rm cm^{-1}}$
for $0.00< \efermi \le 0.40$ eV. The inverse of the phonon life time
is also fitted by $\Gamma_\nu=5-10|\efermi|\ {\rm cm^{-1}}$ for
$-0.20\le \efermi < 0.25$ eV and $\Gamma_\nu=2.5\ {\rm cm^{-1}}$ for
$0.25\le\efermi \le 0.40$ eV so as to reproduce the experimental
results.~\cite{yoon13} It is important to note that the Kohn anomaly
does not give the asymmetric BWF of the $G$ band spectra because the
Kohn anomaly is not an inteference phenomenon; only the interference
effect between the $G$ band and the ERS does however show a BWF
lineshape.  The electron-photon (${\cal M}_{\rm el-op}^{b,a}$) and
electron-phonon (${\cal M}_{{\rm el-}\nu}^{b,a}$) matrix elements for
a transition between states $a \rightarrow b$ are adopted from
previous works within the TB method.~\cite{jiang05, gruneis03} We
approximate the intermediate states (virtual states) to become a real
state with $n=n^\prime$, which is a good approximation for the
resonance condition.~\cite{jiang07b}

The ERS amplitude $A_\ers$ is the summation of the amplitude from the
first-order $A^{(1)}_\ers$ and second-order $A^{(2)}_\ers$
processes. The amplitude of the first-order ERS process is given by
\begin{align}
  A^{(1)}_{\ers}(\omega_{\mathrm{s}}) = & \frac {1}{\pi}
  \sum_{n,n^\prime}\sum_{l,l^\prime} \bigg[ \frac{{\cal M}_{\rm
        el-op}^{n,i}} {\left[\Delta E_{ni}-i\gamma\right]}\nonumber\\&
    \times \frac{K_{n^\prime,l^\prime,n, l}(0)} {\left[\Delta E_{n^\prime
          i}-\hbar\omega_{\rm e}-i(\gamma+\Gamma_{\rm
          e})\right]}\nonumber\\& \times \frac{{\cal M}_{\rm
        el-op}^{f,n^\prime}} {\left[E_{\mathrm{L}}-\hbar\omega_{\rm
          e}-\hbar \omega_\mathrm{s} -i\Gamma_{\rm e}\right]}
    \bigg], \label{eq:ampli1}
\end{align}
where $\omega_{\rm e}$ and $\Gamma_{\rm e}=30$ meV are, respectively,
the energy of the excited DE electron and the inverse life time of the
electron-electron interaction.  The electron-electron interaction
$K_{1,2,3,4}(\qvec)$ defines the scattering of the PE [DE] from an initial
state $(1)\ [(2)]$ to a final state $(3)\ [(4)]$ which consists of
direct $(K^{\rm d})$ and exchange $(K^{\rm x})$ interaction terms,
\begin{equation}
  K_{1,2,3,4}(\qvec) =K^{\rm d}_{1,2,3,4}(\qvec)+K^{\rm
        x}_{1,2,3,4}(\qvec),
\end{equation}
for a spin singlet state.  We do not consider spin triplet states for
simplicity due to the fact that the exchange interaction is
sufficiently small.~\cite{hasdeo13,jiang07b} The direct $K^{\rm
  d}_{1,2,3,4}(\qvec)$ and exchange $K^{\rm x}_{1,2,3,4}(\qvec)$
Coulomb interactions between two electrons in the TB approximation is
given by
\begin{align}
  K_{1,2,3,4}^{\rm d}({\bf q})=& \sum_{ss^\prime={\rm A,B}}C_{s}^1
  C_{s^\prime}^2 C^{*3}_{s} C^{*4}_{s^\prime} \Re \left[v_{ss^\prime}
    \left(\mathbf{q}\right) \right],  \label{eq:Kdfin}\\
  K_{1,2,3,4}^{\rm x}({\bf q})=& \sum_{ss^\prime={\rm A,B}}C_{s}^1
  C_{s^\prime}^2 C_{s^\prime}^{*3} C_{s}^{*4} \Re \left[v_{ss^\prime}
    \left(\mathbf{\kvec^\prime-\kvec-\qvec}\right)
  \right], \label{eq:Kxfin}
\end {align}
where $[1,2,3,4]=[\kvec\cond,\kvec^\prime\val,(\kvec-\qvec)\cond,
(\kvec^\prime+\qvec)\cond]$ in the case of ERS in undoped-graphene
($\efermi = 0$) [Fig.~\ref{Fig2}(a)]. In the electron doped ($\efermi>0$)
and the hole doped ($\efermi<0$) cases, we add possible intraband
transitions $[2,4]=[\kvec^\prime\cond,(\kvec^\prime+\qvec)\cond]$ and
$[2,4]=[\kvec^\prime\val,(\kvec^\prime+\qvec)\val]$, respectively, as
long as state $(2)$ is occupied and state $(4)$ is unoccupied.
$C_{s}^j$ is a tight binding coefficient for an atomic site $s={\rm
  A,\ B}$ and a state $j$.~\cite{sasaki08a} The Fourier transform of
the Coulomb potential $v_{ss^\prime}\left(\mathbf{q}\right)$ is
defined by
\begin{equation}
  v_{ss^\prime}\left(\mathbf{q}\right) = \frac{1}{N} \sum_{u^\prime}
  e^{i\mathbf{q}.\left(\mathbf{R}_{u^\prime s^\prime}-\mathbf{R}_{0s}\right)} v
  \left(\mathbf{R}_{0s},\mathbf{R}_{u^\prime s^\prime}\right),  \label{eq:vq}
\end{equation}
where $v(\Rvec,\Rvec^\prime)$ is the effective Coulomb potential for
the $\pi$ electron system modeled by the Ohno
potential~\cite{perebeinos04,jiang07b}
\begin{equation}
  v\left(\Rvec,\Rvec^\prime\right) =
  \frac{U_{0}}{\sqrt{\left(\frac{4\pi\epsilon_{0}} {e^{2}}U_{0}
        \left|\Rvec-\Rvec^\prime\right|\right)^{2}+1}},
\end{equation}
in which $U_0$ is the on-site Coulomb potential for two $\pi$ electrons at
the same site $\mathbf{R}=\mathbf{R}^\prime$, defined by
\begin{equation}
  \label{eq:onsite}
  U_0=\int d\rvec d\rvec^\prime\frac{e^2}{|\rvec-\rvec^\prime|}
  \phi^2_\pi(\rvec)\phi^2_\pi(\rvec^\prime)
  =11.3\ {\rm eV}.
\end{equation} 
The amplitude of the second-order ERS process is given
by
\begin{align}
  A_{\mathrm{ERS}}(\omega_{\mathrm{s}}) =&
  \frac{1}{\pi}\sum_{n,n^\prime,n^{\prime\prime}}\sum_{m,m^\prime,l,l^\prime}
  \bigg[ \frac{{\cal M}_{\rm el-op}^{n,i}}{\left[\Delta
      E_{ni}-i\gamma\right]} \nonumber\\&\times \frac{K^{\rm
      d}_{n^\prime,m^\prime,n,m}({\bf q})} {\left [\Delta E_{n^\prime
        i}-\hbar\omega_{1} - i(\gamma+\Gamma_{\mathrm{e}}) \right]}
  \nonumber\\& \times \frac{K^{\rm
      d}_{n^{\prime\prime},l^\prime,n^\prime,l}(-{\bf q})}
  {\left[\Delta E_{n^{\prime\prime}
        i}-\hbar\omega_{1}-\hbar\omega_{2}-i(\gamma+
      2\Gamma_{\mathrm{e}})\right]}\nonumber\\
  &\times \frac{{\cal M}_{\rm el-op}^{f,n^\prime}}
  {\left[E_{\mathrm{L}}-\hbar\omega_{1}-\hbar\omega_{2} -
      \hbar\omega_{\mathrm{s}}-2i\Gamma_{\mathrm{e}}\right]} \bigg],
\end{align}
where we also consider the same virtual state approximation as in
Eq.~\eqref{eq:amplitude}.  Here, $\omega_{1}$ and $\omega_{2}$ are the
energies of the DEs emitted for the electron-electron
interaction in the second-order ERS process.  

\section{Electronic Raman Spectra and 
the BWF asymmetry}
\label{sec:ers}

\begin{figure}[t]
\includegraphics[width=8cm]{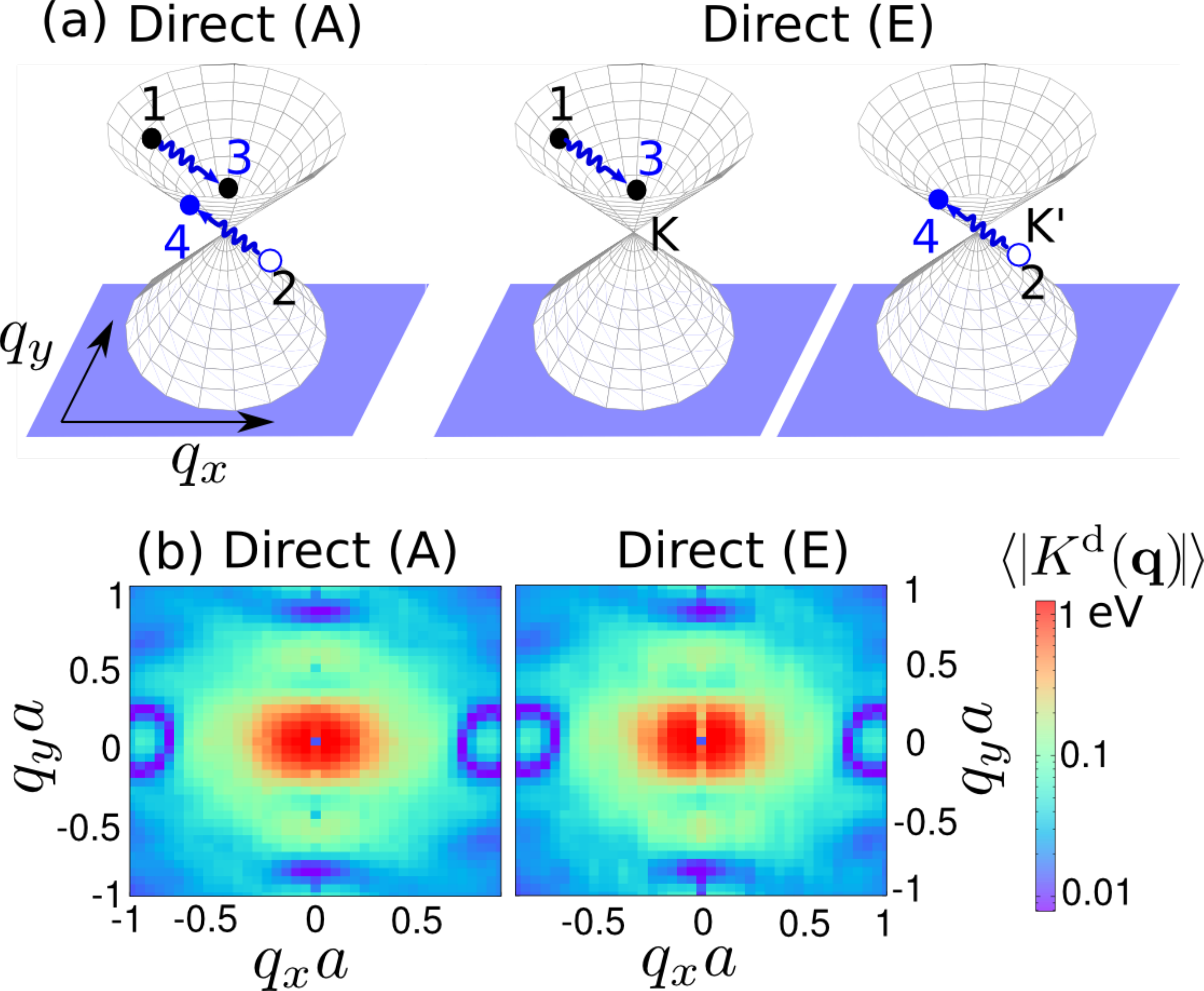}
\caption{\label{Fig2} (Color online) (a) Ilustration of the direct
  Coulomb intravalley (A) interaction and intervalley (E)
  interaction. (b) The averaged absolute value of the direct Coulomb
  interaction matrix element $K^{\rm d}$ as a function of momentum
  transfer $\qvec$ for the intravalley (A) interaction and intervalley
  (E) interaction. The intervalley (e) scattering is not shown in this
  figure.}
\end{figure}

\begin{figure}[t]
\includegraphics[width=80mm]{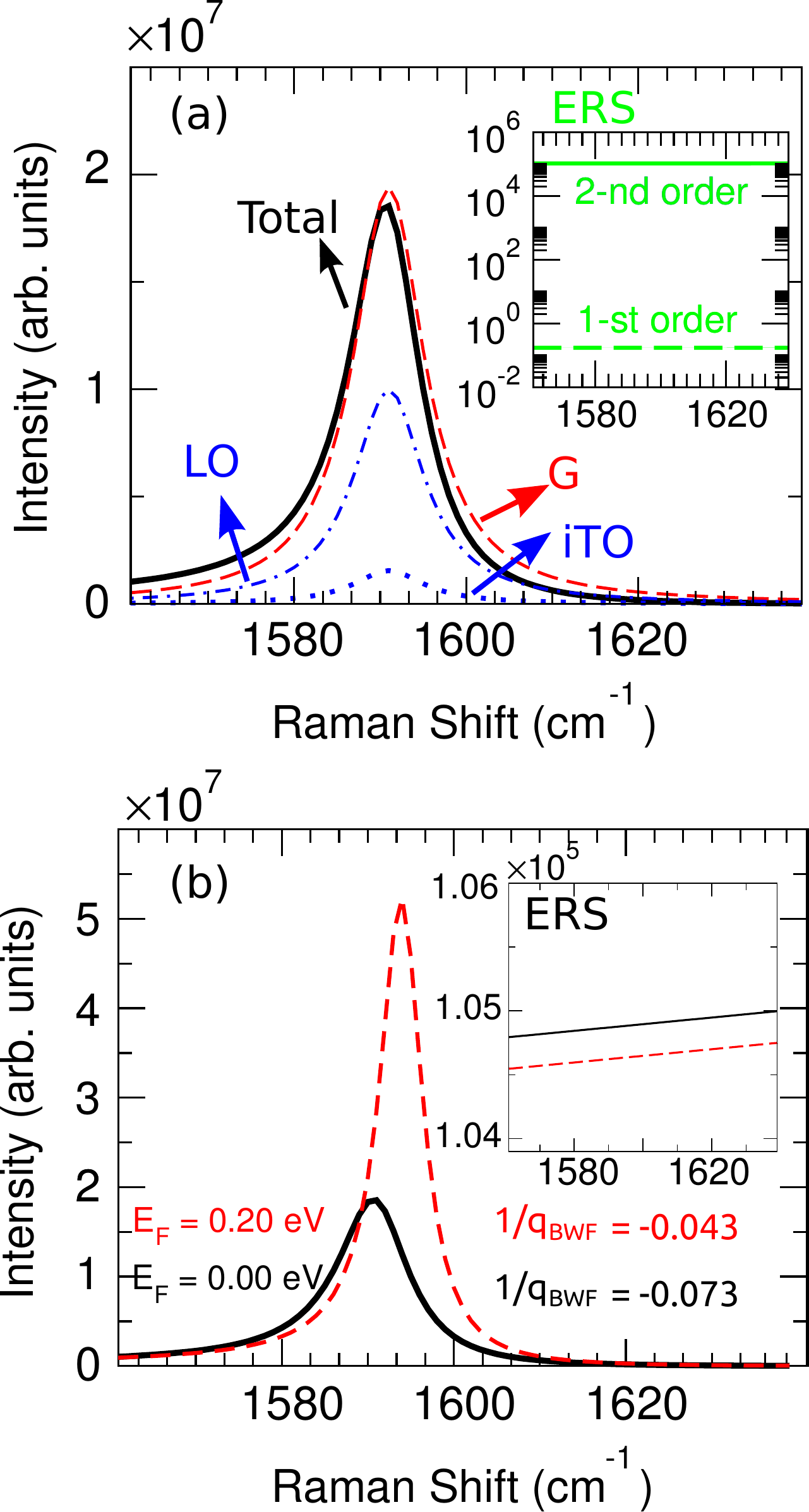}
\caption{\label{Fig3} (Color online) (a) Calculated results of the
  total Raman intensity in Eq.~\eqref{eq:intens} for $\efermi=0.00\
  {\rm eV}$ (solid line) compared with the Lorentzian $G$ mode
  intensity by taking the square of $A_G$ in Eq.~\eqref{eq:amplitude}
  (dashed line).  The $G$ mode constituents, i.e. iTO and LO, are
  indicated by a dotted line and a dot-dashed line, respectively.  An
  asymmetric lineshape (solid line) appears due to the interference
  effect of the $G$ mode with the ERS.  The inset shows calculated
  results of the first-order (dashed line) and the second-order (solid
  line) ERS spectra indicating that the second-order processes have an
  intensity value six-orders of magnitude greater than that of the
  first-order processes. (b) Calculated results of the total Raman intensity for
  $\efermi = 0.00\ {\rm eV}$ (solid line) and $\efermi = 0.20\ {\rm
    eV}$ (dashed line). The BWF asymmetric factor $\qbwf$ decreases by
  increasing the absolute value of $|\efermi|$ away from the Dirac point
  because the ERS intensity also decreases by increasing $|\efermi|$
  (inset).}
\end{figure}

Since the electron-electron interaction depends on initial states
$(1,2)$ of PE and DE and also on a momentum transfer ($\qvec$), we
consider the absolute average value of the matrix elements over the
initial states in order to visualize the strength of the
electron-electron interaction in a simple manner. The direct Coulomb
interaction can occur in either the A or E interaction as shown in
Fig.~\ref{Fig2}(a).  Figure~\ref{Fig2}(b) depicts the absolute average
value of $K^{\rm d}_{1,2,3,4}(\qvec)$ over the initial states $(1,2)$
\begin{equation}
  \langle|K_\mu^{\rm d}(\qvec)|\rangle=\frac{1}{N_1N_2}\sum_{(1,2)}|K^{\rm
    d}_{1,2,3,4}(\qvec)|,
\label{eq:kdavg}
\end{equation}
where $\mu=$ A and E. The e scattering is not shown in
Fig.~\ref{Fig2}(b) for convenient comparison between the A and E
interaction, since the A interaction does not have the e scattering.
$\langle|K_\mu^{\rm d}(\qvec)|\rangle$ only depends on $\qvec$ after
taking the summation over the initial states $(1,2)$ because the final
states $(3,4)$ depend on $(1,2)$ by momentum conservation in
Eq.~\eqref{eq:Kdfin}.  As shown in Fig.~\ref{Fig2}(b), for both the A
and E interactions, $K^{\rm d}$ disappears at $\qvec=0$, indicated by
a small dot at $\qvec=(0,0)$, due to the symmetry of the A and B
sublattice wavefunctions in the graphene unit cell which cancel in
the summation of $\kdir$ in Eq.~\eqref{eq:Kdfin}.~\cite{hasdeo13} The
absence of the direct Coulomb interaction suggests that the ERS should
come from the second order $\qvec\ne0$ electron-electron interaction,
similar to what we found in m-SWNTs.~\cite{hasdeo13} The first-order
ERS can only be possible by means of the exchange Coulomb
interaction. Although we take into account the exchange Coulomb
interaction, the Raman intensity from the first-order process is still
six-orders of magnitude smaller than that of the second order process
[see inset of Fig.~\ref{Fig3}(a)].  Therefore, we can neglect the
first-order processes for both the A and E interactions.

In Fig.~\ref{Fig3} we present the Raman intensity calculation
$I(\omega_{\rm s} )$ of Eq.~\eqref{eq:intens}.  The solid curve in
Fig.~\ref{Fig3}(a) shows the total Raman intensity after considering
the interference of the $G$ mode spectra with the ERS spectra, while
the dashed line shows the Lorentzian $G$ phonon spectra by taking the
square of its probability amplitudes $A_{G}(\omega_{\mathrm{s}})$
[Eq.~\eqref{eq:amplitude}]. The $G$ mode constituents, i.e. the iTO
and LO modes, are indicated by a dotted line and a dot-dashed line,
respectively. It is clear from Fig.~\ref{Fig3}(a) that the calculated
Raman spectra shows asymmetry around the peak position at $1590\ {\rm
  cm^{-1}}$. By fitting the calculated result to Eq.~\eqref{eq:ibwf},
we obtain the fitted values of $\qbwf$, which have the same negative
sign as the experimental data.~\cite{yoon13} For a negative $\qbwf$,
when $\omega_{\rm s}$ is smaller (greater) than $\omega_{\rm G}$,
$I(\omega_{\rm s})$ is greater (smaller) than $|A_G(\omega_{\rm
  s})|^2$, indicating that the interference between the $G$ mode and
the ERS spectra is constructive (destructive) below (above) the
resonance condition $\omega_{\rm s}=\omega_{\rm G}$.

By decreasing (increasing) $\efermi$ further from the Dirac cone,
transitions from (to) the unoccupied (occupied) states are suppressed
due to the Pauli principle. Thus we expect that the asymmetric factor
$\qbwf$ decreases as we change the $\efermi$ from the Dirac point
$\efermi=0.00$ eV to $\efermi=0.20$ eV as shown in
Fig.~\ref{Fig3}(b). The solid line is the intensity of the spectrum
with $\qbwf =-0.073$ when $\efermi=0.00$ eV, while the dashed line is
the corresponding curve with $\qbwf =-0.043$ when $\efermi=0.20$ eV.
The Raman intensity and peak position at $\efermi = 0.20$ eV are larger
than that at $\efermi=0.00$ eV due to the Kohn anomaly effect.~\cite{piscanec04}

Unlike the ERS spectra in m-SWNTs which are Lorentzian
functions,~\cite{farhat11,hasdeo13} the ERS intensity in graphene is a
linear function of $\omega_{\rm s}$ [inset of Fig.~\ref{Fig3}(b)].
The positive gradient of the ERS intensity is due to the greater
scattering path available to excite DEs in the second-order processes
as $\omega_{\rm s}$ increases. The ERS intensity will increase
monotonically and will get saturated at $\omega_{\rm s}\ge\elaser/2$. The
absence of the ERS peak intensity in graphene is related to the
absence of van-Hove singularities within the $G$ mode energy $\sim
0.2$ eV to $\elaser = 2.4$ eV.  The absence of the ERS peak also
becomes the reason why the $\qbwf$ values of the $G$ mode in graphene
are one-order of magnitude smaller compared to that in m-SWNTs.  The
ERS intensity is about two-orders of magnitude smaller than that of the
$G$ mode, and by increasing the $\efermi$ the ERS intensity decreases
only less than $1$\%; nevertheless the change of the $\qbwf$ is
significant [Fig.~\ref{Fig3}(b)].  Thus, this BWF feature is very
sensitive to the presence or absence of the continuum spectra.

\begin{figure*}[t]
\includegraphics[width=16cm]{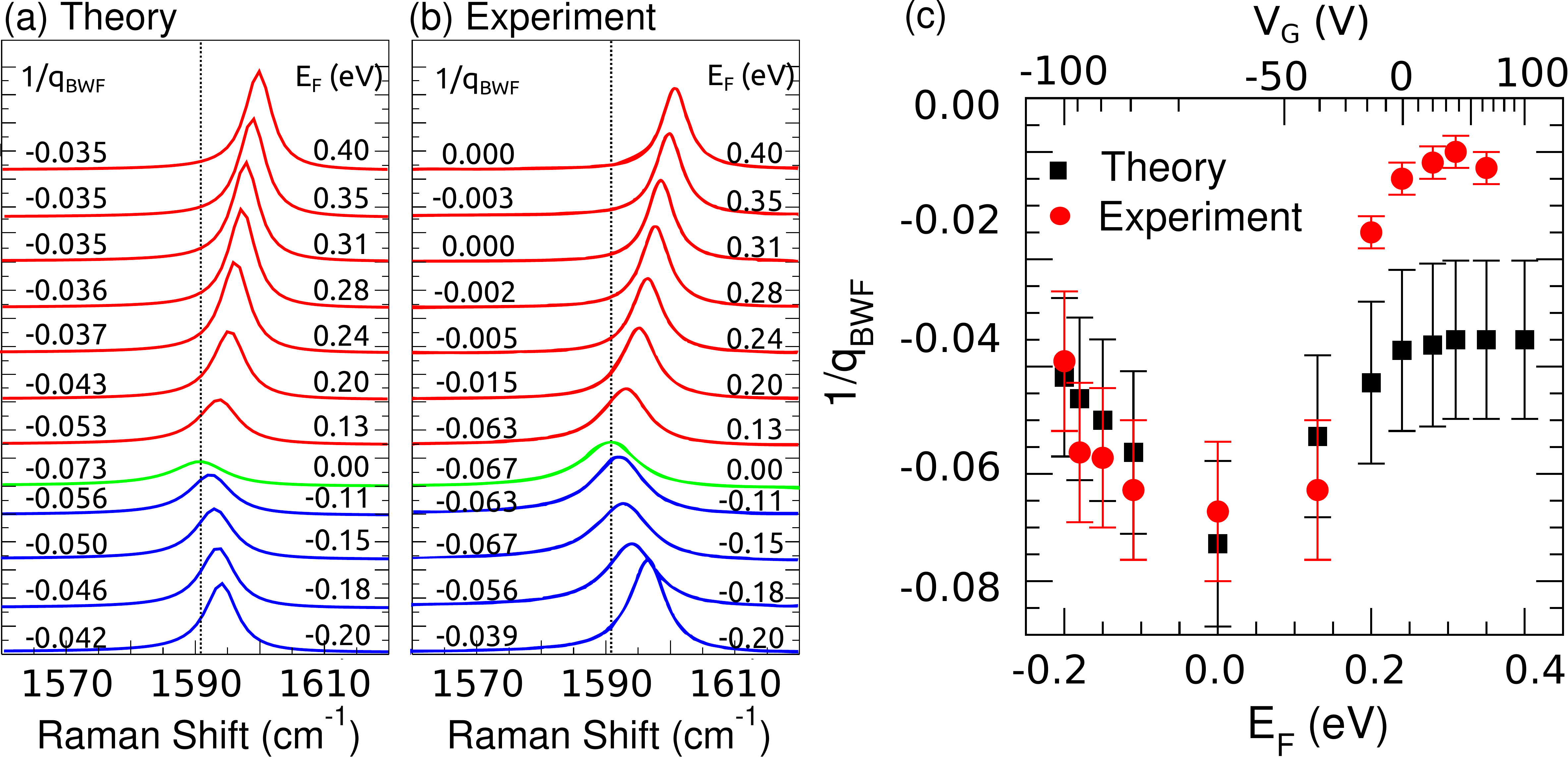}
\caption{\label{Fig4}(Color online) Comparison between (a) the
  calculated results (this work) and (b) the experimental results
  taken from Ref.~\onlinecite{yoon13} for the $G$ band Raman intensity
  as a function of Raman shift, which is plotted for various values of
  $\efermi$ in the range $-0.20\le \efermi \le 0.40$ eV.  The values
  of $\qbwf$ obtained from the calculation and the experiment are also
  given on each plot.  (c) Comparison of the BWF asymmetric factor
  $\qbwf$ as a function of $\efermi$ and gate voltage $V_G$ between
  theory (squares) and experiment (circles).  Both the linewidth and the
  phonon peak frequency-shift due to the Kohn anomaly effect are
  fitted from the experimental results in Ref.~\onlinecite{yoon13}.}
\end{figure*}

In Figures~\ref{Fig4}(a) and (b), we respectively show our calculated
result and corresponding experimental results
(Ref.~\onlinecite{yoon13}) of the $G$ band Raman intensity as a
function of Raman shift, which is plotted for various values of
$\efermi$ in the range $-0.20\le \efermi \le 0.40$ eV.  In the
original version,~\cite{yoon13} Fig.~\ref{Fig4}(b) was given as a
function of gate voltages $V_G$.  For our purpose of comparing the
calculated results and experimental results, here we convert $V_G$ to
$\efermi$ using the relation $\efermi = {\rm sign}(V_G-V_0)\hbar
v_F\sqrt{\alpha \pi |V_G-V_0|}$ where the Fermi velocity $v_F=10^8\
{\rm cm/s}$, the constant voltage adjusted to the Dirac point
$V_0=-57.5\ {\rm V}$, and the capacitance $\alpha = 7.2\times 10^{10}\
{\rm cm^{-2}V^{-1}}$ for the $\rm SiO_2$ dielectric medium with a
thickness $300\ \rm nm$.~\cite{yoon13,saito13,novoselov04} At the
charge neutrality point $\efermi=0.00$ eV, the $G$ band spectrum is
broadened and its frequency is softened due to the Kohn anomaly
effect.  Comparison of the BWF asymmetric factor $\qbwf$ between the
theory (square) and experiment (circle) shows a reasonable agreement
in Fig.~\ref{Fig4}(c) except for $\efermi \ge 0.20$ eV, when the
experimental results deviate from the calculated results.  We suppose
that the deviation is related to the difficulties of observing the BWF
asymmetry at $\efermi>0.20$ eV in the experiment because the continuum
ERS intensity is about two or three-orders of magnitude smaller
compared to the $G$ band intensity.  Such weak ERS spectra might couple
strongly with the background spectra in the experiment which make it
difficult to observe.  The calculated asymmetric factor $\qbwf$ has a
``V''-shaped curve structure as a function of $\efermi$ with the dip
position at $E_F=0.00$ eV.  The decrease of $\qbwf$ is related to the
decrease of the ERS intensity due to the suppression of electron-hole
pair excitations on the Dirac cone upon doping.

The present agreement also reconfirms that plasmons do not contribute
to the continuum spectra.  The reason is as follows.  When $|\efermi|
> 0$, collective excitations (plasmons) are expected to be
generated, and consequently the ERS spectra should be
enchanced.~\cite{hwang07} However, what we obtain in the
present study is that the ERS spectra are in fact suppresed if we
increase $|\efermi|$.  Therefore, we rule out the contribution of
plasmons in the ERS spectra and we conclude that only single-particle
electron-hole pair excitations are important.

\section{Summary} 
\label{sec:summ}
We show that the origin of the BWF spectra in graphene comes from the
continuum single particle electron-hole pair ERS spectra interfering
with the discrete $G$ band phonon spectra.  After calculating the
Raman amplitudes of the ERS and the phonon spectra, we found that
the interference effect between the ERS and the phonon spectra gives a
drastic change in the constructive-destructive interference near the
phonon resonance condition, leading to an asymmetry of the phonon
lineshape when fitted to the BWF lineshape.  By considering the
second-order Raman process, we are able to reproduce the $\efermi$
dependence of the Raman spectra systematically.  We expect that the
asymmetric BWF feature appears generally in the phonon Raman spectra of
all Dirac cone systems.

\section*{Acknowledgments}
The authors are grateful to Duhee Yoon for sharing information
regarding his gate-modulated Raman spectroscopy experiment
(Ref.~\onlinecite{yoon13}). E.H. is supported by a MEXT
scholarship. A.R.T.N. acknowledges a JSPS research fellowship for
young scientists No. 201303921.  R.S.  acknowledges MEXT Grants
No. 25286005 and No. 225107004.  M.S.D. acknowledges NSF-DMR Grant
No. 10-04147.

\bibliographystyle{aip}

\end{document}